\documentclass[12pt]{iopart}

\usepackage{graphicx}
\usepackage{hyperref}
\begin{document}
\title[Field-induced canting in GdCo$_5$ at finite
temperature]{Field-induced canting of magnetic moments in GdCo$_5$ at finite
temperature: first-principles calculations and high-field measurements}
\author{Christopher E.\ Patrick$^1$,
Santosh Kumar$^1$,
Kathrin G\"otze$^1$,
Matthew J.\ Pearce$^1$,
John Singleton$^2$,
George Rowlands$^1$,
Geetha Balakrishnan$^1$,
Martin R.\ Lees$^1$,
Paul A.\ Goddard$^1$,
Julie B. Staunton$^1$}
\address{$^1$Department of Physics, University of Warwick,
Coventry CV4 7AL, UK}
\address{$^2$National High Magnetic Field Laboratory, Los Alamos National Laboratory, 
MS-E536, Los Alamos, New Mexico 87545, USA}
\ead{c.patrick.1@warwick.ac.uk}

\begin{abstract}
We present calculations and experimental measurements of the temperature-dependent magnetization
of a single crystal of GdCo$_5$ in magnetic fields of order 60~T.
At zero temperature the calculations, based on density-functional theory in the disordered-local-moment
picture, predict a field-induced transition from an antiferromagnetic to a canted alignment
of Gd and Co moments at 46.1~T.
At higher temperatures the calculations find this critical field to increase along
with the zero-field magnetization.
The experimental measurements observe this transition to occur between 44--48~T
at 1.4~K.
Up to temperatures of at least 100~K, the experiments continue to observe the transition;
however, at variance with the calculations, no strong temperature
dependence of the critical field is apparent.
We assign this difference to the inaccurate description of the zero-field magnetization
of the calculations at low temperatures, due to the use of classical statistical mechanics.
Correcting for this effect, we recover a consistent description of the high-field
magnetization of GdCo$_5$ from theory and experiment.
\end{abstract}
\maketitle

Of the various families of magnetic intermetallic
compounds formed between rare earths and transition metals (RE-TM), 
the RECo$_5$ series is notable for two reasons.
First, the series includes SmCo$_5$, which 
remains a technologically important permanent magnet thanks to 
its excellent performance at high temperature~\cite{Strnat1967,Gutfleisch2011}.
Second, the relatively simple CaCu$_5$ crystal structure of the RECo$_5$ family~\cite{Kumar1988}
means that experimental observations can often be explained in terms of
a relatively small number of quantities describing fundamental magnetic 
interactions~\cite{Radwanski1986}.
Quantifying these fundamental interactions in RECo$_5$ benefits
the study of a wider range of RE-TM magnets having more diverse crystal 
structures~\cite{Franse1993}.

In general, RE-TM magnets owe many of their excellent properties to the localized
4$f$ electrons of the lanthanide elements~\cite{Elliottbook}.
The 4$f$ electrons can lead to large spin and orbital moments on the RE, while
electrostatic interactions of these electrons with their environment (the crystal field)
can result in a large single-ion magnetocrystalline anisotropy (MCA)~\cite{Kuzmin2008}.
GdCo$_5$ is a rather interesting member of the RECo$_5$ family because,
although each Gd atom carries a large spin moment from having seven unpaired
4$f$ electrons, the spherically-symmetric charge cloud associated with these electrons
causes the orbital moment and crystal field effects to vanish.
Therefore the magnetic anisotropy of GdCo$_5$ is dominated by the MCA 
of the sublattice of Co atoms, with only a small dipolar contribution from the Gd moments.
The magnetic response is determined by a competition between
this MCA, the exchange interaction between the Gd and Co sublattices, and the 
interaction of these sublattices with the external field~\cite{Yermolenko1980}.

The RE-Co exchange interaction in RECo$_5$ is antiferromagnetic, such that in
the absence of external fields the spin moments of the RE and Co atoms align
antiparallel to each other~\cite{Nesbitt1962}.
GdCo$_5$ is therefore a ferrimagnet, whose resultant moment
points in the direction of the Co sublattice moments with a magnitude per formula 
unit (FU) of approximately  (5$\times$1.8 - 7.3 = 1.7)$\mu_\mathrm{B}$~\cite{Patrick2017}.
The exchange field felt by the Gd moments ($\sim$235~T~\cite{Loewenhaupt1994}) is very 
large compared to external fields achievable in the laboratory.
Nonetheless, modest fields of just a few T can break the antiparallel alignment
of Gd and Co moments, provided that the sample is aligned with its magnetic easy axis
(the crystallographic $c$ direction) pointing normal to the applied field~\cite{Ballou1986}.
If instead the external field is applied along the easy axis, the antiparallel
alignment is expected to be stable up to at least 70~T~\cite{Radwanski1992}.

An intermediate case has the GdCo$_5$ sample free to rotate
in the applied field, which is the natural geometry for powder samples~\cite{Liu1994}.
Radwa\'nski et al.~\cite{Radwanski1992} predicted that canting between the Gd and Co moments 
would be induced in this setup with an applied field of 40~T, resulting in an abrupt
change of gradient in the magnetization vs field curve.
Subsequent measurements in pulsed fields of 60~T reported by Kuz'min et al.~\cite{Kuzmin2004}
confirmed the existence of this feature, which occurred at 46~T at 5~K.
More recent work by Isnard et al.~\cite{Isnard2012} on another Gd-Co compound, GdCo$_{12}$B$_6$,
was able to demonstrate both the transition from the antiparallel to the canted state,
and also the corresponding feature at much higher field marking the transition to parallel (ferromagnetic)
Gd-Co alignment.
The same work reported magnetization curves at different temperatures up to 121~K,
allowing the temperature dependence of the exchange coupling to be investigated~\cite{Isnard2012}.

Although models of ferrimagnets in external fields have been developed at least
as far back as 1968~\cite{Clark1968}, these models generally require experimental parameterization.
However, some of us~\cite{Patrick2018} recently introduced a method of calculating temperature-dependent 
magnetization versus field curves from first principles (FPMVB).
We developed the method, which is based on relativistic density-functional theory in the disordered-local-moment
picture (DFT-DLM)~\cite{Gyorffy1985,Staunton2006}, in order to understand magnetization measurements on GdCo$_5$ orientated with its
easy axis normal to a relatively small  ($\leq$7~T) applied field.
The purpose of this Letter is to show how the same FPMVB approach can be applied to a free-to-rotate
sample in much higher ($\leq 60$~T) fields.
We pair our calculations with pulsed-field measurements of a single crystal of GdCo$_5$ in temperatures
up to 100~K.
These new measurements allow us to compare the theoretical and experimental values of the temperature-dependent critical
fields required to induce the transition from the antiparallel to the canted Gd-Co sublattices in GdCo$_5$.

\begin{figure}
\includegraphics[width=83mm]{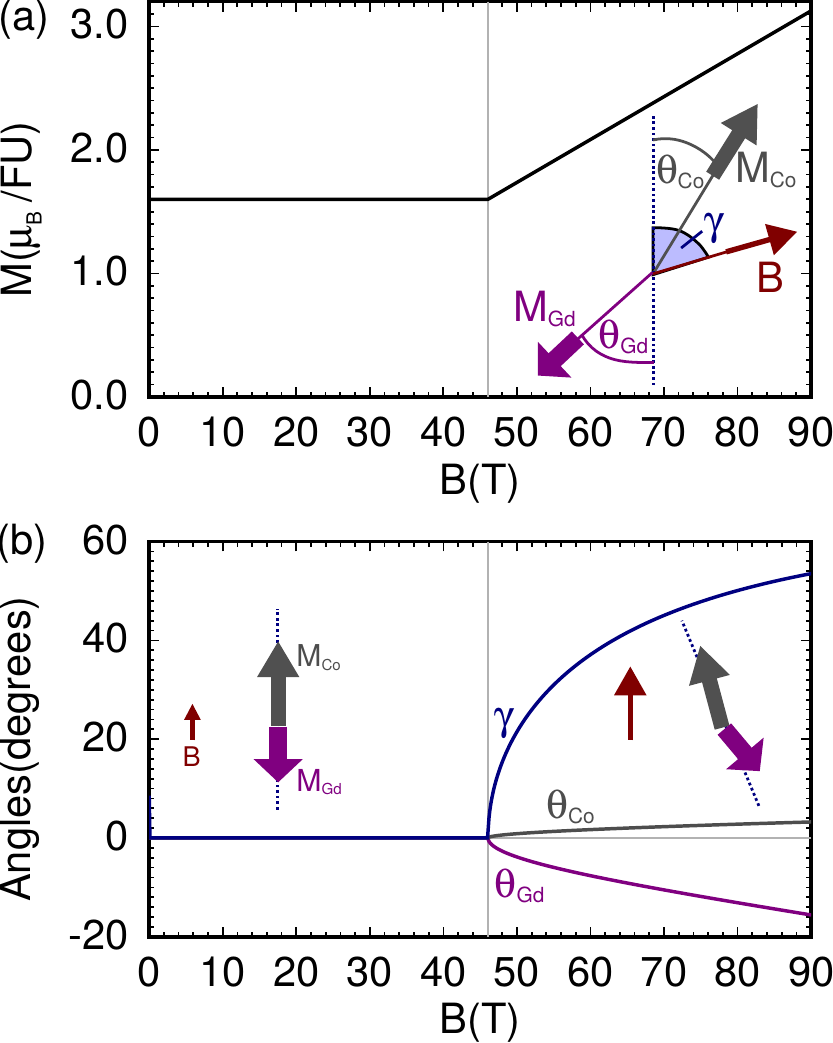}
\caption{
(a) Magnetization vs field curve calculated at $T=0$~K 
for a GdCo$_5$ sample free to rotate in the applied field.
The inset illustrates the angles used in the definition of 
$F^\mathrm{Tot}_2$, with the dotted line representing the crystallographic
$c$ axis.
The vertical grey line marks the transition from antiparallel to canted 
magnetic sublattices.
(b) Calculated variation of the angles $\{\theta_\mathrm{Gd},\theta_\mathrm{Co},\gamma\}$ 
at $T=0$~K.
The arrows illustrate the orientations of the magnetic sublattices with respect
to the applied field.
 \label{fig.1}}
\end{figure}

We first discuss our calculations of the magnetization versus field curves.
In our previous work~\cite{Patrick2018} we showed how these curves could be calculated either
by carefully analyzing the torque on each magnetic moment, or by parametrizing
a model expression for the free energy which could then be minimized for a given external field.
Here we take the second approach,
using the model that we previously found to give 
an accurate description of the free energy landscape~\cite{Patrick2018}:
\begin{eqnarray}
F_2(\theta_\mathrm{Gd},\theta_\mathrm{Co})
&=& - A \ \mathrm{cos}(\theta_\mathrm{Gd}-\theta_\mathrm{Co}) + 
K_\mathrm{1,Co} \ \mathrm{sin}^2\theta_\mathrm{Co} \nonumber \\
&&+ K_\mathrm{2,Co} \ \mathrm{sin}^4\theta_\mathrm{Co}
+ K_\mathrm{1,Gd} \ \mathrm{sin}^2\theta_\mathrm{Gd}  \nonumber \\
&& + S(\theta_\mathrm{Gd},\theta_\mathrm{Co}).
\end{eqnarray}
The angles  $\theta_\mathrm{Gd}$ and $\theta_\mathrm{Co}$ are given
with respect to the crystallographic $c$ axis, as shown in the inset
of Figure~\ref{fig.1}(a).
The quantities $A$, $K_{i,\mathrm{X}}$ and $S$ are all dependent on temperature.
$A$ describes the antiferromagnetic Gd-Co exchange, while
the various  $K_{i,\mathrm{X}}$ values quantify the MCA of the individual sublattices 
originating from the spin-orbit interaction.
Note that although the Gd-4$f$ electrons do not contribute to the MCA, 
the Gd-5$d$ electrons do result in a small positive value for
$K_\mathrm{1,Gd}$.
$S(\theta_\mathrm{Gd},\theta_\mathrm{Co})$ is the contribution to the MCA from dipole-dipole interactions, 
and has the explicit form~\cite{Ballou1987}:
\begin{eqnarray}
S(\theta_\mathrm{Gd},\theta_\mathrm{Co}) &=& S_1 \ \mathrm{sin}^2\theta_\mathrm{Gd}
+ S_2 \ \mathrm{sin}^2\theta_\mathrm{Co} + S_3 \times \nonumber \\
&&
\left( \sin\theta_\mathrm{Gd} \sin\theta_\mathrm{Co}
- \frac{2}{3} \cos(\theta_\mathrm{Gd}-\theta_\mathrm{Co})\right). \nonumber
\end{eqnarray}

The additional contribution to the free energy due to an external field $\vec{B}$ is $-\sum_i \vec{B}\cdot\vec{M_i}$, where
$\vec{M_\mathrm{Gd}}$ and  $\vec{M_\mathrm{Co}}$ are the magnetizations of the two sublattices.
While $M_\mathrm{Gd}$ is independent of magnetization direction, 
the calculations showed
a temperature-dependent magnetization anisotropy on the Co
sublattice~\cite{Patrick2018}, well described by the expression 
$M_\mathrm{Co}(\theta_\mathrm{Co}) = M^0_\mathrm{Co} [ 1 - p \sin^2\theta_\mathrm{Co}]$.
For a field applied at an angle $\gamma$ to the $c$ axis,
our expression for the free energy is therefore
\begin{eqnarray}
F^\mathrm{Tot}_2(\theta_\mathrm{Gd},\theta_\mathrm{Co},\gamma,B)
&=& F_2(\theta_\mathrm{Gd},\theta_\mathrm{Co}) \nonumber \\
&&+ BM_\mathrm{Gd}\cos(\theta_\mathrm{Gd}-\gamma) \nonumber \\
&&- BM^0_\mathrm{Co} \cos(\theta_\mathrm{Co} - \gamma)\times \nonumber \\
&& [ 1 - p \sin^2\theta_\mathrm{Co}].
\label{eq.Ft2}
\end{eqnarray}
Assuming that the GdCo$_5$ sample is able to reach its equilibrium state, the magnetization
measured along the field direction $\sum_i \vec{\hat{B}}\cdot\vec{M_i}$ is determined by the set of angles
$\{\theta_\mathrm{Gd},\theta_\mathrm{Co},\gamma\}$ which minimize $F^\mathrm{Tot}_2$.
In the case that the sample is clamped, $\gamma$ is fixed according to the experimental geometry.

We emphasize that the sublattice magnetizations $M_\mathrm{Co}$ and $M_\mathrm{Gd}$
depend on the temperature $T$.
In the disordered-local-moment picture of magnetism, $\vec{M_\mathrm{X}} = \mu_\mathrm{X} \vec{m_\mathrm{X} }(T)$,
where $\mu_\mathrm{X}$ is the local moment magnitude (e.g.\ $\sim7\mu_\mathrm{B}$ for Gd) and $\vec{m_\mathrm{X}}$
is an order parameter quantifying the configurationally-averaged orientation of the local moment.
The magnitude of $\vec{m_\mathrm{X}}$ varies from one at 0~K to zero at the Curie temperature $T_\mathrm{C}$.
We note that, in principle, $\vec{m_\mathrm{X}}$ should also depend on external field, 
e.g.\ to describe paramagnetic behaviour.
However, given the high $T_\mathrm{C}$ of GdCo$_5$ ($\sim1000$~K), up to room temperature we expect the external field to 
have a minor effect on the order parameter compared to thermal fluctuations.
Therefore, in our calculations we always 
use the zero-field values of $m_\mathrm{X}$.

We performed the numerical minimization of $F^\mathrm{Tot}_2$ using the standard
Broyden-Fletcher-Goldfarb-Shanno scheme as implemented in the \texttt{SciPy} distribution~\cite{SciPy}.
The temperature-dependent quantities $A$, $K_{i,\mathrm{X}}$ etc.\ required to construct 
$F^\mathrm{Tot}_2$ were calculated from first principles in our previous paper~\cite{Patrick2018} and
are reproduced in the Supplementary Material.

In Figure~\ref{fig.1}(a) we show the magnetization versus field
curve calculated by  minimizing $F^\mathrm{Tot}_2$ at $T = 0$~K.
As in experiment, the magnetization is measured along the field direction.
Up to 46.1~T the applied field does not induce any change in the magnetization,
which remains at the value $M^0_\mathrm{Co} - M_\mathrm{Gd}$ = 1.60$\mu_\mathrm{B}$/FU.
At this critical value $B_\mathrm{C}$, there is a kink 
in the magnetization curve. 
Above $B_\mathrm{C}$ the measured magnetization becomes effectively 
linear with respect to the applied field, and
at much higher fields (463~T, not shown) the magnetization saturates
to $M^0_\mathrm{Co} + M_\mathrm{Gd}$.

Further insight into the behaviour of the magnetization can be obtained
by plotting the field dependence of the magnetization angles
$\theta_\mathrm{Gd}$ and $\theta_\mathrm{Co}$, and of the angle 
$\gamma$ between the easy $c$ axis and the applied field [Figure~\ref{fig.1}(b)].
At $B_\mathrm{C}$ the sample undergoes a sudden rotation (increase in $\gamma$)
accompanied by a rotation of the sublattice magnetizations away from the easy axis.
The Co sublattice rotates by a relatively small amount compared to the Gd sublattice
(1.7$^\circ$ compared to 7.3$^\circ$ at 60~T, respectively).
As indicated by the schematics in Figure~\ref{fig.1}(b), the effect
of these rotations is to reduce the misalignment of the Gd magnetization
with the external field.
Precisely how this misalignment is reduced is a balance of the energy penalties
associated with breaking the antiparallel alignment of the Co and Gd sublattices,
of misaligning the Co magnetization with the external field, and 
of rotating the individual sublattice magnetizations away from the easy axis.
As we show below, this latter anisotropy penalty is weak compared
to the exchange and external contributions.

\begin{figure}
\includegraphics[width=83mm]{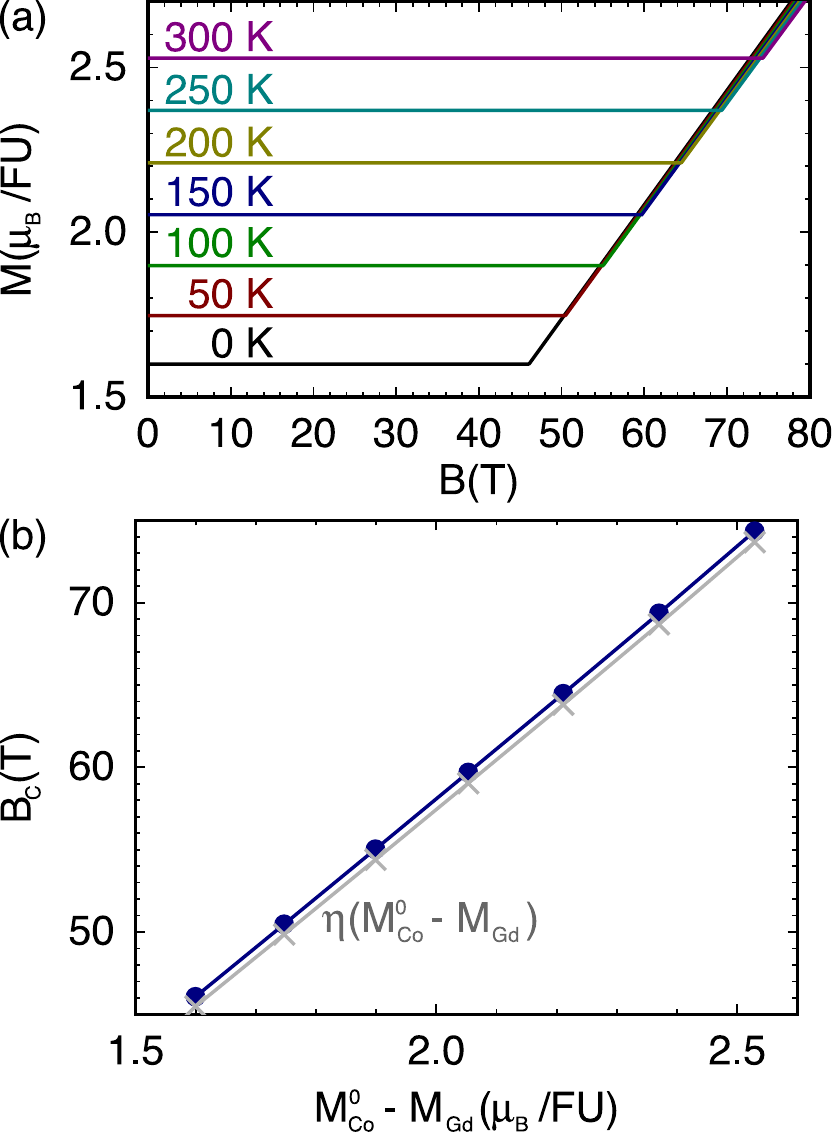}
\caption{
(a) Magnetization vs field curves calculated using
$F^\mathrm{Tot}_2$ and temperature-dependent parameters.
(b) Critical magnetic fields $B_\mathrm{C}$ at which the magnetization
displays a kink, extracted from the curves of (a) (blue circles)
or calculated using the expression $\eta(M^0_\mathrm{Co} - M_\mathrm{Gd})$
(grey crosses), where $\eta \equiv A/(M^0_\mathrm{Co}M_\mathrm{Gd})$ (see text).
The lines joining the symbols are guides to the eye.
\label{fig.2}}
\end{figure}

In Figure~\ref{fig.2}(a) we show the magnetization curves calculated using
$F^\mathrm{Tot}_2$ and the temperature-dependent parameters, up to 300~K.
The increasing zero-field magnetization as a function of temperature
is a consequence of the ferrimagnetic nature of GdCo$_5$, where
the Gd moments disorder more quickly with temperature than Co~\cite{Patrick2017}.
Accordingly, the resultant $M^0_\mathrm{Co} - M_\mathrm{Gd}$ initially increases with
temperature, although as we discuss later the calculated rate of increase exceeds
what is observed experimentally.
All of the curves have the same qualitative form as the $T = 0$~K case, and
the magnetizations for $B > B_\mathrm{C}$ lie almost on top of each other.

Like the zero-field magnetization, $B_\mathrm{C}$ increases monotonically with temperature.
In Figure~\ref{fig.2}(b) we plot $B_\mathrm{C}$ as a function of   
$M^0_\mathrm{Co} - M_\mathrm{Gd}$ to show that there is effectively a linear relation
between the two quantities.
To understand this behaviour further it is useful to consider a simpler two-sublattice
model~\cite{Radwanski1986}, where the free energy is modelled as 
\begin{eqnarray}
F^\mathrm{tot}_3 &=& - A \ \mathrm{cos}(\theta_\mathrm{Gd}-\theta_\mathrm{Co}) + 
K_\mathrm{1,Co} \ \mathrm{sin}^2\theta_\mathrm{Co} \nonumber \\
&&+ B [M_\mathrm{Gd}\cos(\theta_\mathrm{Gd}-\gamma) - M^0_\mathrm{Co} \cos(\theta_\mathrm{Co} - \gamma)].
\end{eqnarray}
Minimizing this model expression analytically yields three simple results: first, that for $K_\mathrm{1,Co} \geq 0$
the Co moments always lie along the easy axis, i.e.\ $\theta_\mathrm{Co} = 0$ or 180$^\circ$; second, that 
$B_\mathrm{C} = \eta [M^0_\mathrm{Co}-M_\mathrm{Gd}  ]$
where $\eta \equiv A / (M_\mathrm{Gd} M^0_\mathrm{Co})$; and third, that the measured magnetization above
$B_\mathrm{C}$ is given by $M(B) = B/\eta$, until the upper critical field of $\eta [M^0_\mathrm{Co} + M_\mathrm{Gd} ]$ is reached.

We recalculated the magnetization vs field curves using  $F^\mathrm{tot}_3$ and obtained
results that, on the scale of Figure~\ref{fig.2}(a), are indistinguishable from
those obtained from $F^\mathrm{tot}_2$.
In Figure~\ref{fig.2}(b) we plot $B_\mathrm{C}$ predicted from $F^\mathrm{tot}_3$ as
grey crosses.
The two sets of $B_\mathrm{C}$ closely resemble each other, with the plot scale obscuring the variation in the 
offset of 0.59--0.67~T.
Part of this offset is due to the dipolar anisotropy $S_3$ renormalizing $A$, and the rest due
to the anisotropy energy of the Gd sublattice.

As indicated by the near-linearity of the crosses in Fig.~\ref{fig.2}(b), our calculated
values of $\eta$ depend only weakly on temperature, varying from 28.4--29.5~T/($\mu_\mathrm{B}$/FU)
over the 0--300~K range.
This small variation is consistent with the high-field study on 
the related material GdCo$_{12}$B$_6$~\cite{Isnard2012}, which
could not resolve any temperature variation of $\eta$ between 4.2~and~63~K.

Our calculated value of 46.1~T for $B_\mathrm{C}$ at zero temperature agrees rather well with
the value of 46~T measured at 5~K by Kuz'min et al.~\cite{Kuzmin2004}.
However, to our knowledge measurements on GdCo$_5$ at higher temperatures
have not been reported in the literature.
Therefore, using the single crystal of GdCo$_5$ grown by 
some of us recently using the floating zone technique~\cite{Patrick2017}, 
we carried out
high-field measurements at temperatures between 1.4~and~100~K.
The pulsed-field magnetization measurements were performed at 
the National High Magnetic Field Laboratory in Los Alamos; 
fields of up to 60~T with typical rise times $\approx$ 10~ms were used.
The single crystal, free to rotate, is placed in a 1.3~mm diameter 
polychlorotrifluoroethylene ampoule (inner diameter 1.0~mm) that 
can be moved into and out of a 1500-turn, 1.5~mm bore, 1.5~mm long 
compensated-coil susceptometer, constructed from 50 gauge high-purity 
copper wire~\cite{Goddard2008}.
When the sample is within the coil and the field pulsed the voltage 
induced in the coil is proportional to the rate of change of 
magnetization with time.
Accurate values of the magnetization are obtained by numerical 
integration of the signal with respect to time, followed by 
subtraction of the integrated signal recorded using an empty coil 
under the same conditions.
The magnetic field is measured via the signal induced within a 
coaxial 10-turn coil and calibrated via observation of de~Haas-van Alphen 
oscillations arising from the copper coils of the susceptometer.

\begin{figure}
\includegraphics{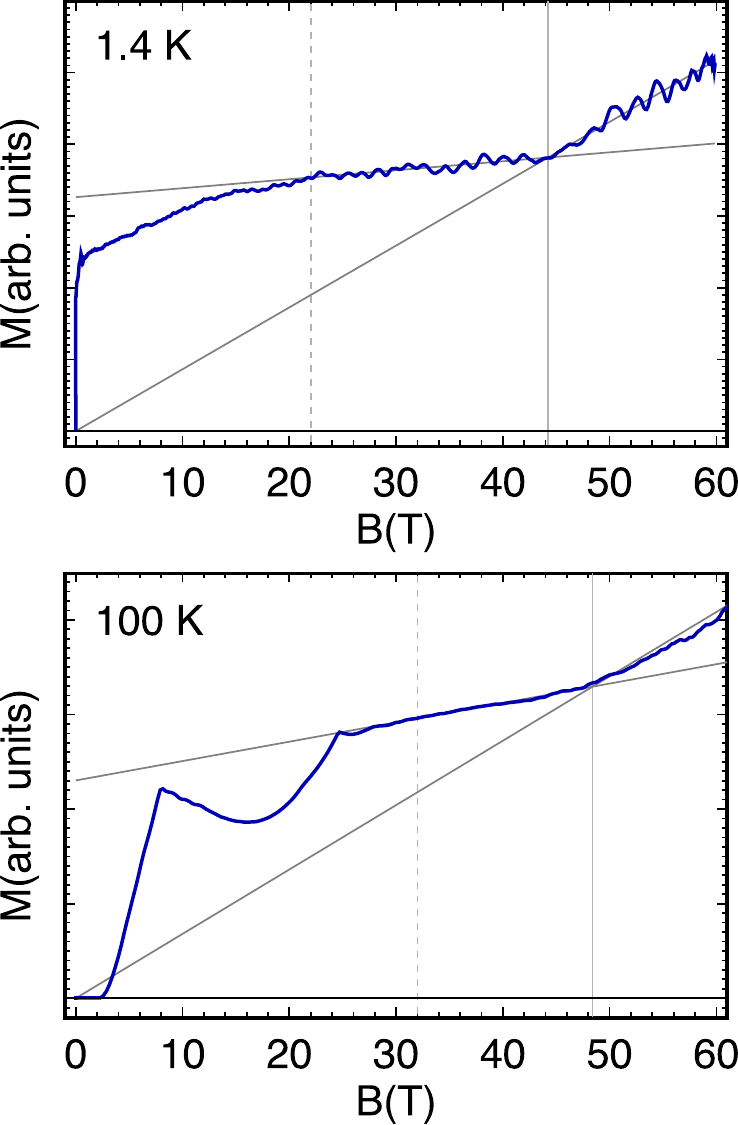}
\caption{Experimentally-measured magnetization vs field
curves at 1.24~and~100~K (blue).  The horizontal black line
is the zero axis; the $y$-axis scale is arbitrary.
The grey lines are explained in the text. \label{fig.3}}
\end{figure}

Figure~\ref{fig.3} shows the experimental magnetization vs field
curves measured in a pulsed field of 60~T, at 1.4~K and 100~K.
The experimental curves show more features than
the idealized calculations.
In particular, since the sample is initially demagnetized and randomly oriented
as the field is switched on, it undergoes a variable
amount of motion which results in large fluctuations in the low-field
magnetization.
Similarly as the field is reduced to zero the sample demagnetizes, which
is not accounted for in the calculations.
In general the experiments do not show a region of strictly constant
magnetization unlike the calculations, which neglect the field dependence of
the order parameter as discussed above.
However, the experimental curves do exhibit the same critical behaviour
at high fields, namely a kink in the magnetization curve.
At 1.4~K this kink occurs at 44~T, reasonably close to the previously reported
value of 46~T at 5~K~\cite{Kuzmin2004}.

In order to perform a quantitative comparison between calculations and
experiment up to 100~K, we developed a protocol to extract $B_\mathrm{C}$ from
the experimental data in a consistent way.
The protocol consists of first separating each
set of data into two curves, corresponding to an increasing and decreasing magnetic field, and analyzing
them separately.
For instance, in Figure~\ref{fig.3} the 1.4~K and 100~K data were taken for decreasing and increasing field,
respectively.
Next we discard the data at low fields where the signal is dominated by the sample motion and (de)magnetization.
The rejected data falls to the left of the dashed vertical lines in Figure~\ref{fig.3}.
We then partition the remaining data into two regions and fit the data within each region with a straight line.
The partitioning is performed under the constraints that 
(a) the two fitted lines join at the partition and (b) the second line intersects the origin.
Constraint (b) is guided by the numerical results of Figures~\ref{fig.1}~and~\ref{fig.2} which 
showed the high-field magnetization to effectively satisfy this constraint (the relation is exact
for $F^\mathrm{Tot}_3$).
The straight lines fits are shown in grey in Figure~\ref{fig.3}.
The field at which the two lines join (i.e.\ the partition) is taken as the experimental $B_\mathrm{C}$, indicated by
the solid vertical line.

\begin{figure}
\includegraphics[width=83mm]{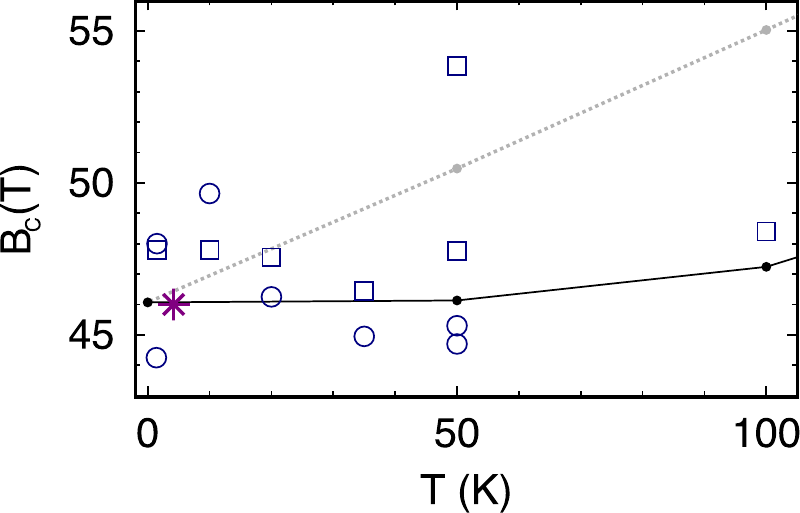}
\caption{
Critical magnetic fields $B_\mathrm{C}$ extracted from our magnetization curves, either
for increasing (squares) or decreasing (open circles) applied field.
The star is the previously reported value of 46~T~\cite{Kuzmin2004}.
The filled circles joined by straight lines are the calculated values of $B_\mathrm{C}$, with
the grey values corresponding to Figure~\ref{fig.2}(b) and the black values corrected
for the experimental magnetization, as discussed in the text.
\label{fig.4}}
\end{figure}

We applied the described protocol to all of the magnetization curves measured at different temperatures.
In all but two cases the procedure gave unambiguous values for $B_\mathrm{C}$, which we plot in Figure~\ref{fig.4}.
All of the measured curves, including the two failed cases, are shown in the Supplementary Material.

Apart from an anomalous value of 54~T at 50~K,
the experimental values of $B_\mathrm{C}$ lie in the 44--50~T region.
For the same pulse there is a difference of 1--2~T depending on whether the increasing or decreasing
applied field is analyzed (squares or circles in Figure~\ref{fig.4}).
However, at variance with the calculated values of $B_\mathrm{C}$ (grey circles and dotted lines)
the experiments do not show any particular increase in critical field with temperature. 

To resolve this apparent discrepancy we return to Figure~\ref{fig.2}(b),
which shows how $B_\mathrm{C}$ is essentially linear in the zero-field magnetization
$M^0 = M^0_\mathrm{Co} - M_\mathrm{Gd}$.
Unfortunately, the DFT-DLM theory used to calculate the parameters determining the free energy
does not describe the temperature dependence of the magnetization very well at low temperatures.
This is because the DFT-DLM statistical mechanics is based on a classical Heisenberg ($J=\infty$) description
of spins~\cite{Gyorffy1985}, with no barrier to rotating the spins at zero temperature.
So, while the calculations predict the GdCo$_5$ zero-field magnetization to increase
by 0.3$\mu_\mathrm{B}$/FU between 0--100~K, experimentally the increase is in fact just 0.04$\mu_B$/FU~\cite{Patrick2017}.

To correct for this effect, we rescale the 
DFT-DLM critical fields as 
$B^\mathrm{corr}_\mathrm{C}(T) = B_\mathrm{C}(0) + \Delta B_\mathrm{C}(T) \times \Delta M_\mathrm{exp}^0(T) / \Delta M_\mathrm{calc}^0(T)$,
where $\Delta M^0$ denotes the change in zero-field magnetization with temperature, either measured
experimentally or calculated.
$B_\mathrm{C}(T)$ are the uncorrected critical fields, with  $\Delta B_\mathrm{C}(T) = B_\mathrm{C}(T) - B_\mathrm{C}(0)$.
Plotting the rescaled fields as the black circles in Figure~\ref{fig.4} we see how, 
even with the temperature dependence of $\eta$ included,
the calculated change in  $B^\mathrm{corr}_\mathrm{C}$ over the 0--100~K temperature range
is now only of the order of 1~T, which is below the resolution of the experiment.
Therefore our interpretation is that the lack of variation in $B_\mathrm{C}$ observed experimentally 
is consistent with the theory, once the low temperature behaviour of the DFT-DLM calculations has been accounted for.
The experimental observations are also consistent with the previously-reported measurements on 
GdCo$_{12}$B$_6$~\cite{Isnard2012} which found a similarly small variation in critical field with temperature.

In summary, we have demonstrated how the FPMVB method, developed and parametrized to 
calculate magnetization curves at low applied fields, can also be used to calculate
high-field behaviour.
Close to zero temperature, there is good agreement between the predictions of the theory
and experiment.
At higher temperatures, the qualitative predictions of the model remain accurate
to at least 100~K, where we continued to observe the predicted kink in the magnetization
curve; however, to obtain quantitative agreement in the temperature dependence of
the critical magnetic field it is necessary to correct for the classical statistical mechanics
employed in the calculations.
Interesting avenues for future exploration include more complicated magnetic
systems consisting of multiple magnetic sublattices, as well as systems where
competing anisotropies lead to unusual magnetization behaviour, e.g.\ first-order
magnetization processes~\cite{Thang1996}.

\section*{Acknowledgments}
The present work forms part of the PRETAMAG project,
funded by the UK Engineering and Physical Sciences Research
Council, Grant No. EP/M028941/1.
Crystal growth work at Warwick is also supported by 
EPSRC Grant no.\ EP/M028771/1.
The work has also received funding from the European Research Council (ERC) under 
the European Union's Horizon 2020 research and innovation 
programme (Grant agreement No. 681260).
A portion of this work was performed at the National High Magnetic Field Laboratory, 
which is supported by National Science Foundation Cooperative Agreement No.\
DMR-1157490, the State of Florida, and the US Department of Energy (DoE) 
and through the DoE Basic Energy Science Field Work Proposal ``Science in 100 T''.

\section*{References}

\providecommand{\newblock}{}


\begin{thebibliography}{10}
\expandafter\ifx\csname url\endcsname\relax
  \def\url#1{{\tt #1}}\fi
\expandafter\ifx\csname urlprefix\endcsname\relax\def\urlprefix{URL }\fi
\providecommand{\eprint}[2][]{\url{#2}}

\bibitem{Strnat1967}
Strnat K, Hoffer G, Olson J, Ostertag W and Becker J~J 1967 {\em J. Appl.
  Phys.\/} {\bf 38} 1001

\bibitem{Gutfleisch2011}
Gutfleisch O, Willard M~A, Br\"uck E, Chen C~H, Sankar S~G and Liu J~P 2011
  {\em Adv. Mater.\/} {\bf 23} 821

\bibitem{Kumar1988}
Kumar K 1988 {\em J. Appl. Phys.\/} {\bf 63} R13

\bibitem{Radwanski1986}
Radwa\'nski R 1986 {\em Physica B+C\/} {\bf 142} 57

\bibitem{Franse1993}
Franse J~J~M and Radwa\'nski R~J 1993 {\em Handbook of Magnetic Materials\/}
  vol~7 (Elsevier North-Holland, New York) chap~5, p 307

\bibitem{Elliottbook}
Elliott R~J 1972 {\em Magnetic Properties of Rare Earth Metals\/} ed Elliott
  R~J (London and New York: Plenum Press) p~1

\bibitem{Kuzmin2008}
Kuz'min M~D and Tishin A~M 2008 {\em Handbook of Magnetic Materials\/} vol~17
  (Elsevier B.V.) chap~3, p 149

\bibitem{Yermolenko1980}
Yermolenko A~S 1980 {\em Fiz. Metal. Metalloved.\/} {\bf 50} 741

\bibitem{Nesbitt1962}
Nesbitt E~A, Williams H~J, Wernick J~H and Sherwood R~C 1962 {\em J. Appl.
  Phys.\/} {\bf 33} 1674

\bibitem{Patrick2017}
Patrick C~E, Kumar S, Balakrishnan G, Edwards R~S, Lees M~R, Mendive-Tapia E,
  Petit L and Staunton J~B 2017 {\em Phys. Rev. Materials\/} {\bf 1} 024411

\bibitem{Loewenhaupt1994}
Loewenhaupt M, Tils P, Buschow K and Eccleston R 1994 {\em J. Magn. Magn.
  Mater.\/} {\bf 138} 52

\bibitem{Ballou1986}
Ballou R, D\'eportes J, Gorges B, Lemaire R and Ousset J 1986 {\em J. Magn.
  Magn. Mater.\/} {\bf 54} 465

\bibitem{Radwanski1992}
Radwa\'nski R, Franse J, Quang P and Kayzel F 1992 {\em J. Magn. Magn.
  Mater.\/} {\bf 104} 1321

\bibitem{Liu1994}
Liu J, de~Boer F, de~Ch\^atel P, Coehoorn R and Buschow K 1994 {\em J. Magn.
  Magn. Mater.\/} {\bf 132} 159

\bibitem{Kuzmin2004}
Kuz'min M~D, Skourski Y, Eckert D, Richter M, M\"uller K~H, Skokov K~P and
  Tereshina I~S 2004 {\em Phys. Rev. B\/} {\bf 70} 172412

\bibitem{Isnard2012}
Isnard O, Skourski Y, Diop L~V~B, Arnold Z, Andreev A~V, Wosnitza J, Iwasa A,
  Kondo A, Matsuo A and Kindo K 2012 {\em J. Appl. Phys.\/} {\bf 111} 093916

\bibitem{Clark1968}
Clark A~E and Callen E 1968 {\em J. Appl. Phys.\/} {\bf 39} 5972

\bibitem{Patrick2018}
Patrick C~E, Kumar S, Balakrishnan G, Edwards R~S, Lees M~R, Petit L and
  Staunton J~B 2018 {\em Phys. Rev. Lett.\/} {\bf 120} 097202

\bibitem{Gyorffy1985}
Gy\"orffy B~L, Pindor A~J, Staunton J, Stocks G~M and Winter H 1985 {\em J.
  Phys. F: Met. Phys.\/} {\bf 15} 1337

\bibitem{Staunton2006}
Staunton J~B, Szunyogh L, Buruzs A, Gy\"orffy B~L, Ostanin S and Udvardi L 2006
  {\em Phys. Rev. B\/} {\bf 74} 144411

\bibitem{Ballou1987}
Ballou R, D\'eportes J and Lemaire J 1987 {\em J. Magn. Magn. Mater.\/} {\bf
  70} 306

\bibitem{SciPy}
Jones E, Oliphant T, Peterson P {\em et~al.\/} 2001-- {SciPy}: Open source
  scientific tools for {Python} \urlprefix\url{http://www.scipy.org/}

\bibitem{Goddard2008}
Goddard P~A, Singleton J, Sengupta P, McDonald R~D, Lancaster T, Blundell S~J,
  Pratt F~L, Cox S, Harrison N, Manson J~L, Southerland H~I and Schlueter J~A
  2008 {\em New J. Phys.\/} {\bf 10} 083025

\bibitem{Thang1996}
Thang C, Brommer P, Colpa J, Thuy N and Franse J 1996 {\em Physica B: Condensed
  Matter\/} {\bf 228} 205

\end{thebibliography}
\end{document}